# The Luminosity Distribution of Short Gamma-Ray Bursts under a Structured Jet Scenario

Qi Guo[1,2], Daming Wei[1,2], and Yuanzhu Wang[1]
[1] Key Laboratory of Dark Matter and Space Astronomy, Purple Mountain Observatory, Chinese Academy of Sciences, Nanjing 210034, People's Republic of China
dmwei@pmo.ac.cn
[2] School of Astronomy and Space Science, University of Science and Technology of China, Hefei, Anhui 230026, People's Republic of China
*Received 2019 April 26; revised 2020 March 5; accepted 2020 March 28; published 2020 April 30*

## Abstract

The joint detection of gravitational wave (GW) and electromagnetic radiation from the binary neutron star merger event GW170817 marks a breakthrough in the field of multi-messenger astronomy. The short gamma-ray burst (sGRB) GRB 170817A, associated with this binary neutron star merger event, has an isotropic-equivalent gamma-ray radiation luminosity of $1.6 \times 10^{47}$ erg s$^{-1}$, which is much lower than that of other sGRBs. The measurement of the superluminal movement of the radio afterglow emission confirms the presence of the relativistic jet, and the emission features can be well explained by the structured jet model. In this paper, we calculate the luminosity distribution of sGRBs and its evolution with redshift based on the structured (Gaussian) jet model, and find that the typical luminosity increase with redshift, for nearby sGRBs (such as for luminosity distance less than 200 Mpc) the typical gamma-ray luminosity is just around $10^{47}$–$10^{48}$ erg s$^{-1}$, which naturally explains the very low radiation luminosity of GRB 170817A. We derived the detection probability of sGRBs by Fermi-GBM and found that the expected detection rate of sGRBs is only about 1 yr$^{-1}$ within the distance of several hundred Mpc. We explored the effect of the power-law index $\alpha$ of the merger time distribution on the observed characteristics and found that it had little effect on the observed luminosity and viewing-angle distributions. However, it is very interesting that, for different values of $\alpha$, the distributions of the number of observed sGRBs are quite different, so it is possible to determine the value of $\alpha$ through observed distributions of the number of sGRBs. We used the Bayesian method to make a quantitative analysis and found that the value of $\alpha$ may be identified when the number of observed sGRBs with known redshifts is more than 200. Finally, we compare our results of gamma-ray luminosity distribution with sGRBs with known redshifts, and found that our results are consistent with the observation, which implies that our simulation results can reproduce the observed luminosity distribution well.

*Unified Astronomy Thesaurus concepts:* Gamma-ray bursts (629); Gamma-ray astronomy (628); Astronomical methods (1043)

## 1. Introduction

Gamma-ray bursts are the most violent explosions in the universe. Based on their durations, gamma-ray bursts are usually classified into two types, long gamma-ray bursts (LGRBs) and short gamma-ray bursts (sGRBs), which have different origins (for reviews, see Mészáros 2006; Zhang 2007; Gehrels et al. 2009). LGRBs are usually thought to originate from the collapse of massive stars (Woosley 1993). For a long time, sGRBs were only associated by indirect evidence with the merging of double compact stars (Nakar 2007); the indirect evidence included the location of sGRBs in host galaxies, non-detection of associated supernova (SN), large galaxy offsets, weak spatial correlation between sGRBs and star formation regions within their host galaxies (Nakar 2007; Berger 2014, for reviews), and, in particular, the identification of the so-called Li-Paczynski macronovae/kilonova in GRB 130603B (Berger et al. 2013; Tanvir et al. 2013), GRB 060614 (Jin et al. 2015; Yang et al. 2015), GRB 050709 (Jin et al. 2016) and recently in GRB 070809 (Jin et al. 2020). The joint detection of gravitational wave (GW) and gamma-ray emission from the binary neutron star merger event GW170817/GRB 170817A marks a breakthrough in the field of multi-messenger astronomy, and provides the most conclusive evidence that at least a fraction of sGRBs are indeed from the merger of double neutron stars (Abbott et al. 2017a, 2017b; Goldstein et al. 2017; Savchenko et al. 2017).

However, for GRB 170817A, the first sGRB associated with a GW signal, its prompt gamma-ray and afterglow emission show some peculiar characteristics, which are different from other sGRBs, for example, the isotropic-equivalent prompt gamma-ray emission energy is just about $3 \times 10^{46}$ erg and its luminosity is just about $1.6 \times 10^{47}$ erg s$^{-1}$, much lower than that of typical sGRBs. In addition, the X-ray and radio afterglows first rose for several months and then declined quickly. Although the early afterglow emission could not differentiate between a cocoon model or a structured jet model, the late afterglow features strongly favor the structured jet model with a large viewing angle (e.g., Lazzati et al. 2018; Troja et al. 2019), and the presence of a relativistic jet with a large viewing angle has been identified by the successful measurement of the superluminal movement of the radio afterglow emission (Mooley et al. 2018; Ghirlanda et al. 2019).

In the standard fireball model, the ejecta is assumed to be a conical outflow within which the energy density and Lorentz factor are constant (the so-called "top-hat" jet), while, beyond the ejecta, the energy density and the Lorentz factor reduce to zero abruptly (Pian et al. 2017). But in reality, the energy density and the Lorentz factor should vary with the polar angle, which has been confirmed by some numerical simulations (Aloy et al. 2005; Murguia-Berthier et al. 2017). In previous studies, structured jet models have been proposed to investigate the characteristics of the prompt and afterglow emission of LGRBs (Granot et al. 2002; Rossi et al. 2002; Zhang & Mészáros 2002; Wei & Jin 2003), while for sGRBs it is hard to infer the energy profile of the jet since the observational data are usually rare, the exception is GRB 051221A, for which the X-ray afterglow light curves showed a flat segment and can be well accounted for by a two-component jet model (Jin et al. 2007).

Motivated by the discovery of GW170817 and GRB 170817A association, it is reasonable to assume that the sGRBs driven by binary neutron star mergers would have similar jet structure and may be observed with relative large viewing angles, so it is important to study the characteristics of these possible GW-associated sGRBs. In this paper, we investigate the observable features (such as the distribution of luminosity and viewing angles) of these off-axis sGRBs that may be associated with GW events using the structured jet model.

The paper is organized as follows. In Section 2 we describe the method we used to study the observable characteristics of nearby sGRBs, including the rate of sGRBs, the luminosity distribution and the Gaussian jet we adopted. Our results are presented in Section 3 and conclusions are given in Section 4.

## 2. The Method

When calculating the emission from a GRB jet, the most widely used jet model is a "top-hat" jet, however, the "top-hat" jet cannot explain the emission of GRB 170817A, while the structured jet model is strongly favored. For structured jet models, there is usually a critical angle $\theta_c$ (the jet's core). The energy and the Lorentz factor distribution are nearly independent on the polar angle $\theta$ when $\theta < \theta_c$, while for $\theta > \theta_c$, the energy and the Lorentz factor drop quickly. When a GRB is viewed at an angle larger than the jet's core ($\theta > \theta_c$, e.g., "off-axis"), the GRB's emission would be weak and the afterglow would rise first and then decline quickly, just as the case of GRB 170817A. The emission received by off-axis observers strongly depends on the energy density and Lorentz factor distribution within the jet. In the literature, three types of structured jet models are usually adopted: (a) the power-law distribution model $\epsilon(\theta) = \epsilon_0 \theta^{-k}$ for $\theta > \theta_c$ and $\epsilon(\theta) = \epsilon_0$ for $\theta < \theta_c$ (Dai & Gou 2001; Rossi et al. 2002; Zhang & Mészáros 2002; Wei & Jin 2003); (b) the Gaussian-type jet $\epsilon(\theta) = \epsilon_0 \exp(-\theta^2/\theta_c^2)$ (Zhang et al. 2004), for $\theta_c \sim 5°$; and (c) the two-component jet model (Berger et al. 2003; Wu et al. 2005; Huang et al. 2006). Here we use the Gaussian distribution to describe the energy and Lorentz factor distribution (e.g., Zhang & Mészáros 2002; Beniamini & Nakar 2019)

$$L(\theta) = \frac{dE}{dtd\Omega} = \frac{L_0}{4\pi} e^{-(\theta^2/\theta_c^2)} \quad (1)$$

$$\Gamma(\theta) = 1 + (\Gamma_0 - 1) e^{-(\theta^2/\theta_c^2)}. \quad (2)$$

Here, $L(\theta)$ is the luminosity per solid angle in the observer frame (Howell et al. 2019), $\Gamma(\theta)$ is the Lorentz factor, $\theta_c$ represents the jet's core, and $L_0$ is the isotropic-equivalent luminosity observed on-axis. The observed gamma-ray luminosity for an observer located at angle $\theta_{\rm obs}$ from the jet's axis is given by (Kathirgamaraju et al. 2018; Beniamini & Nakar 2019)

$$L_{r,\rm obs}(\theta_{\rm obs}) = \eta_\gamma \int \frac{L(\theta)}{\Gamma(\theta)} \delta^3(\theta, \phi, \theta_{\rm obs}) d\Omega \quad (3)$$

$$\delta(\theta, \phi, \theta_{\rm obs}) = \frac{1}{\Gamma(\theta)(1 - \beta(\theta)\cos\chi)} \quad (4)$$

$$\cos\chi = \cos\theta_{\rm obs}\cos\theta + \sin\theta\cos\phi \quad (5)$$

where $\delta$ is the Doppler factor, $\beta(\theta)$ is the velocity corresponding to $\Gamma(\theta)$, $\chi$ is the angle between the emitting material and the observer, and $\eta_\gamma$ is the efficiency of conversion from the kinetic energy to the $\gamma$-rays. It is easy to show that, for $\theta_{\rm obs} \ll \theta_c$, the emission is dominated by "line of sight" emitters and the observed isotropic-equivalent $\gamma$-ray luminosity is just $\eta_\gamma \cdot L_0$, while, for larger viewing angles, the emission is dominated by "off line of sight" emitters and the isotropic-equivalent gamma-ray luminosity would be much lower (Beniamini & Nakar 2019).

In order to study the characteristics of sGRBs, we also need to know the formation rate and luminosity function of sGRBs, which are discussed in the following subsections.

### 2.1. The Rate of sGRBs

It has been widely believed that sGRBs originate from the merging of binary compact objects, i.e., neutron star–neutron star (NS–NS) or neutron star–black hole (NS–BH) coalescence, and this has been confirmed by the discovery of the GW170817–GRB 170817A association. In this merging scenario, the formation rate of sGRBs can be written as (Ando 2004)

$$R_{\rm sGRB}(t) \propto \int_{t_F}^{t} dt' R_{\rm SF}(t') P_m(t - t') \quad (6)$$

where $R_{\rm SF}(t)$ is the star formation rate (SFR), $P_m(t)$ is the probability distribution function of merging time of the binary system from its formation, $t_F$ represents the formation epoch of galaxies, and, in many papers, $z(t_F) = 5$ is usually adopted (Ando 2004). Although there remain a huge amount of uncertainties concerning the cosmic SFR history, especially in the high-redshift universe, the SFR in the low-redshift region is fairly well known by many observations at various wave bands. Among many models of SFR, two models are often used.

1. Madau & Dickinson (Madau & Dickinson 2014)

$$R_{\rm SF1}(z) = 0.015 \frac{(1+z)^{2.7}}{1 + [(1+z)/2.9]^{5.6}} M_\odot \ {\rm yr}^{-1} {\rm Mpc}^{-3}. \quad (7)$$

2. Porciani (Porciani & Madau 2001)

$$R_{\rm SF2}(z) = 0.16 h_{70} \frac{\exp(3.4z)}{\exp(3.4z) + 22}$$
$$\times \frac{[\Omega_m(1+z)^3 + \Omega_\Lambda]^{1/2}}{(1+z)^{3/2}} M_\odot \ {\rm yr}^{-1} {\rm Mpc}^{-3}. \quad (8)$$

Figure 1 shows the SFR history for the above two models. It is obvious that the behaviors of these two models are roughly the same in the low-redshift region $z < 1.5$, while, at $z > 1.5$, they are significantly different. For the function SF1, it gives an exponentially decreasing formation rate, while, for the function SF2 it gives the constant formation rate at $z > 1.5$. In this paper, we use the latest model of star formation rate (SF1) to calculate the distribution of sGRBs, and adopt the standard ∧CDM cosmology ($\Omega_m = 0.3$, $\Omega_\wedge = 0.7$, $H_0 = 70 h_{70}$ km s$^{-1}$ Mpc$^{-3}$).

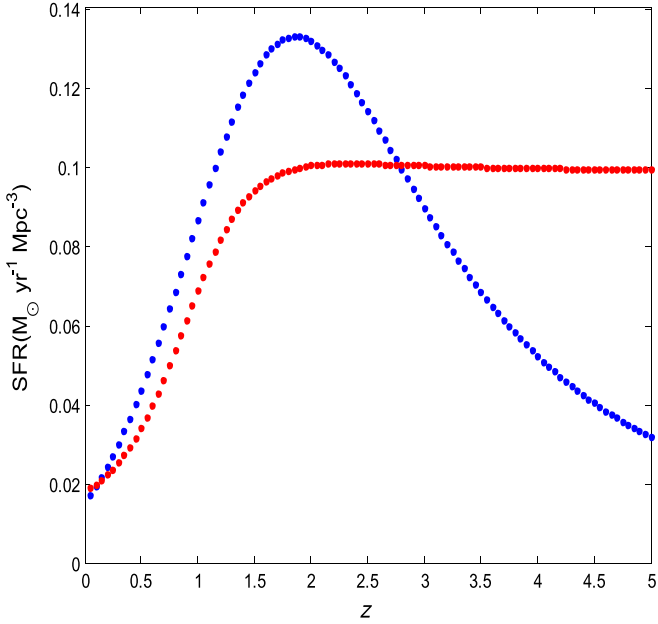

**Figure 1.** Two models of star formation rate. The blue and the red lines represent the models "Madau & Dickinson" and "Porciani," respectively.

In addition to the SFR, the merger time distribution is also essential to estimate the sGRB rate. In previous works, a simple parameterization function such as $Pm(t) \propto t^\alpha$ with a lower cutoff timescale $\tau$ is usually adopted (Porciani & Madau 2001; Schmidt et al. 2001a, 2001b; Ghirlanda et al. 2004), and, in general, the cutoff time $\tau_m = 20$ Myr is adopted (Bulik et al. 2004; Regimbau & Hughes 2009; Meacher et al. 2015), where $\tau_m$ represents the lower cutoff timescale in units of Myr. Previous studies have shown that the distribution of the merging times depends on the distribution of the initial orbital separation modeled as $dN/da \propto a^{-\beta_N}$, and the expected merger times follow $dN/dt_{\rm merge} \propto \tau^{-\alpha}$, where $\alpha \equiv -\beta_N/4 - 3/4$. One can see that, even if $\beta_N$ varies in an extreme range from 0 to 7, $\alpha$ is between 0.75 and 2.5 (Ando 2004; Belczynski et al. 2017, 2018). In order to examine whether the value of $\alpha$ would influence the sGRB distribution significantly, we take the values of $\alpha$ as $(-0.5, -1.0, -1.5, -2.5)$ to calculate the sGRB rate according to Equation (6), adopting the rate of local NS–NS mergers to be 110–3840 Gpc$^{-3}$ yr$^{-1}$ (Abbott et al. 2019). The results are shown in Figure 2 and we find that the different values of index $\alpha$ can make the sGRB rate quite different.

### 2.2. The Luminosity Function of sGRB

When calculating the luminosity of sGRBs, we need to know the distribution of the isotropic-equivalent luminosity $L_0$. In GRB studies, the broken power-law model is most commonly used as the form of the luminosity function so, in this paper, we take the standard broken power-law distribution as follows

$$\Phi(L_0) = \Phi_0 \begin{cases} \left(\dfrac{L_0}{L_*}\right)^{-\alpha_L} & L_0 < L_*, \\ \left(\dfrac{L_0}{L_*}\right)^{-\beta_L} & L_* \leqslant L_0. \end{cases} \quad (9)$$

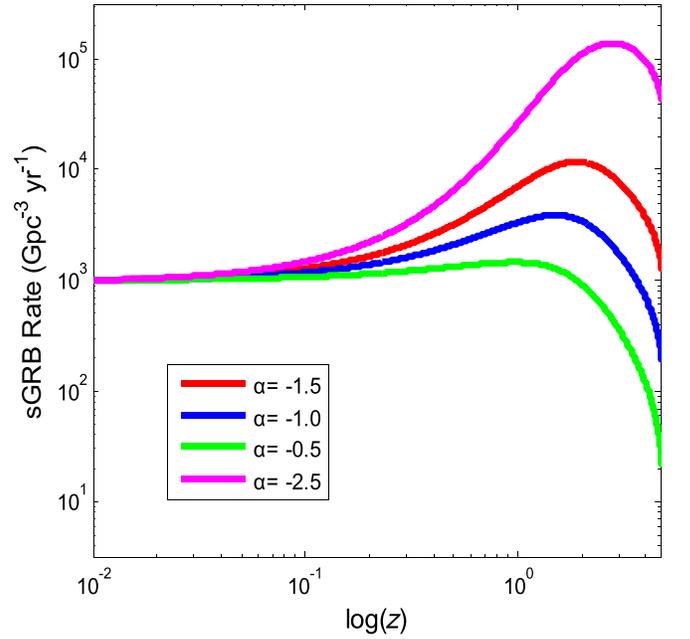

**Figure 2.** The formation rate history of sGRBs that is obtained with the assumption that they are associated with binary neutron star mergers. The rate of local NS–NS mergers is taken to be $\rho_0 = 1000$ Gpc$^{-3}$ yr$^{-1}$ (Abbott et al. 2019). The green, blue, red, and purple lines correspond to different values of $\alpha$ with $\alpha = -0.5, -1.0, -1.5, -2.5$.

Here $L_*$ is the critical luminosity and $\alpha_L$ and $\beta_L$ are the slopes describing the low and high component of the luminosity function, respectively. Following the paper of Wanderman & Piran (2015), we take the values $\alpha_L = 1$, $\beta_L = 2$, and $L_* = 2 \times 10^{52}$ erg s$^{-1}$.

### 2.3. k-correction

The luminosity of sGRBs can be written as $L = 4\pi d_L^2 k(z) F_0$, where $F_0$ is the observed energy flux of the sGRBs in units of erg cm$^{-2}$ s$^{-1}$, $k(z)$ is the cosmological k-correction factor (see Bloom et al. 2001)

$$K = \frac{\int_{1\,\rm keV}^{10^4\,\rm keV} E f(E) dE}{\int_{E_{\rm min}(1+z)}^{E_{\rm max}(1+z)} E f(E) dE}, \quad (10)$$

$$f(E) = \begin{cases} A\left(\dfrac{E}{100\text{ keV}}\right)^{\alpha_*} \times \exp\left(-\dfrac{(2+\alpha_*)E}{E_p}\right) & E < \dfrac{(\alpha_* - \beta_*)E_p}{2+\alpha_*} \\ A\left(\dfrac{E}{100\text{ keV}}\right)^{\beta_*} \times \exp^{\beta_* - \alpha_*}\left(\dfrac{(\alpha_* - \beta_*)E_p}{(2+\alpha_*)100\text{ keV}}\right)^{\alpha_* - \beta_*} & E \geqslant \dfrac{(\alpha_* - \beta_*)E_p}{2+\alpha_*} \end{cases} \quad (11)$$

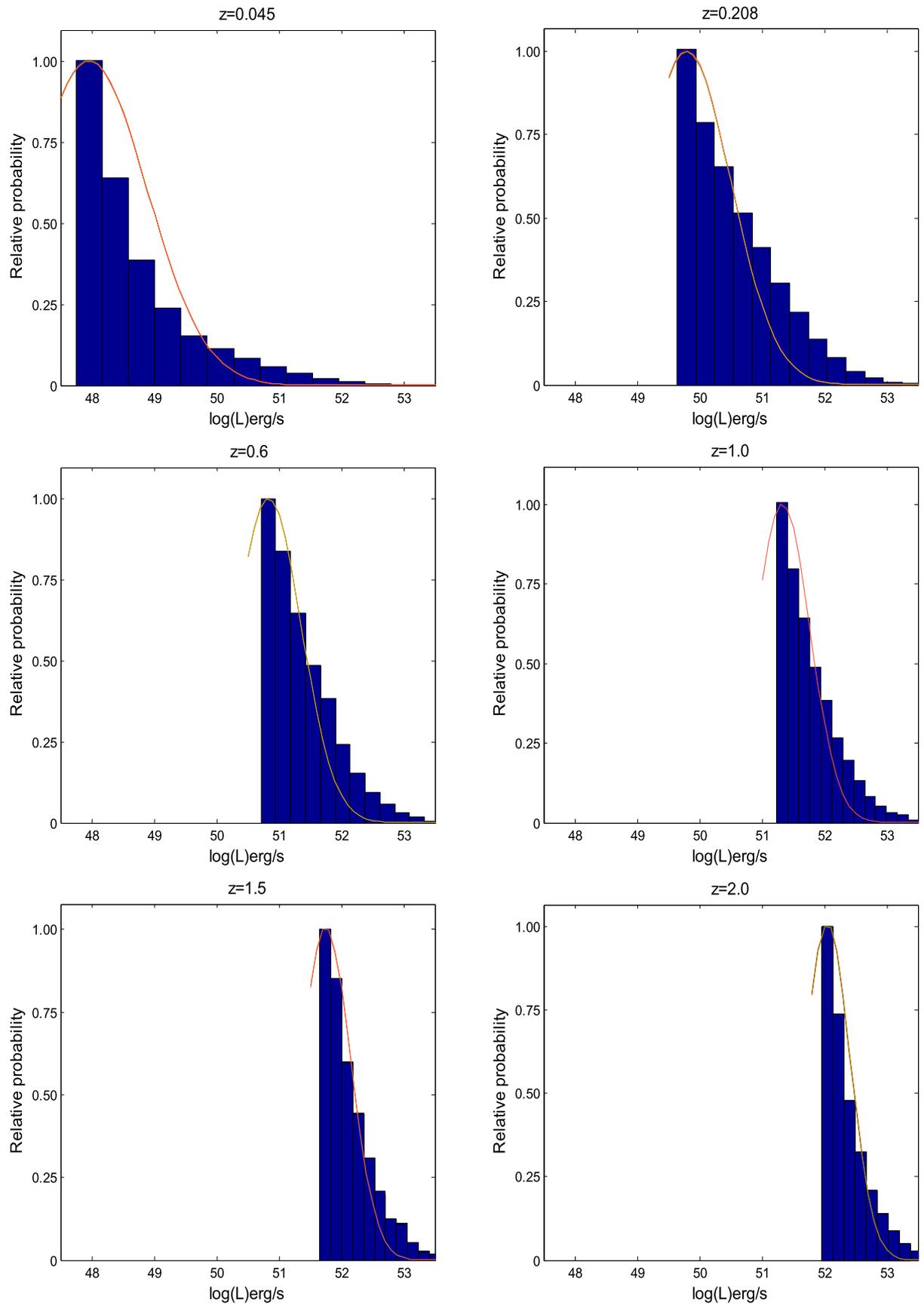

**Figure 3.** The distribution of luminosity at different redshifts. The dotted line is the Gaussian fitting of the distribution. We also plot the lines for different values of $\alpha$ (from $-0.5$ to $-2.5$), and these lines are nearly undistinguishable.

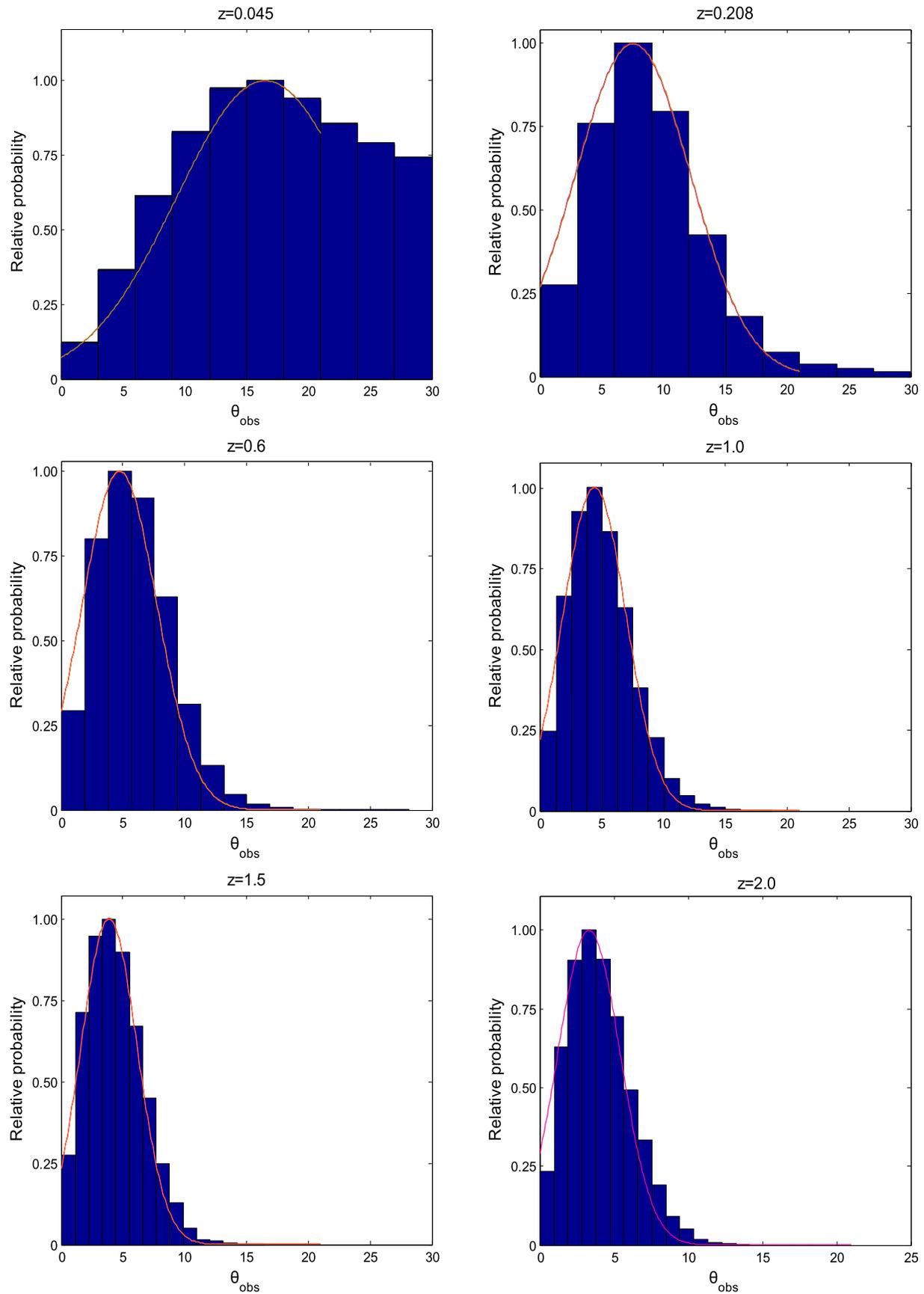

**Figure 4.** The distribution of viewing angle at different redshifts. The dotted line is the Gaussian fitting of the distribution. We also plot the lines for different values of $\alpha$ (from $-0.5$ to $-2.5$), and these lines are nearly undistinguishable.

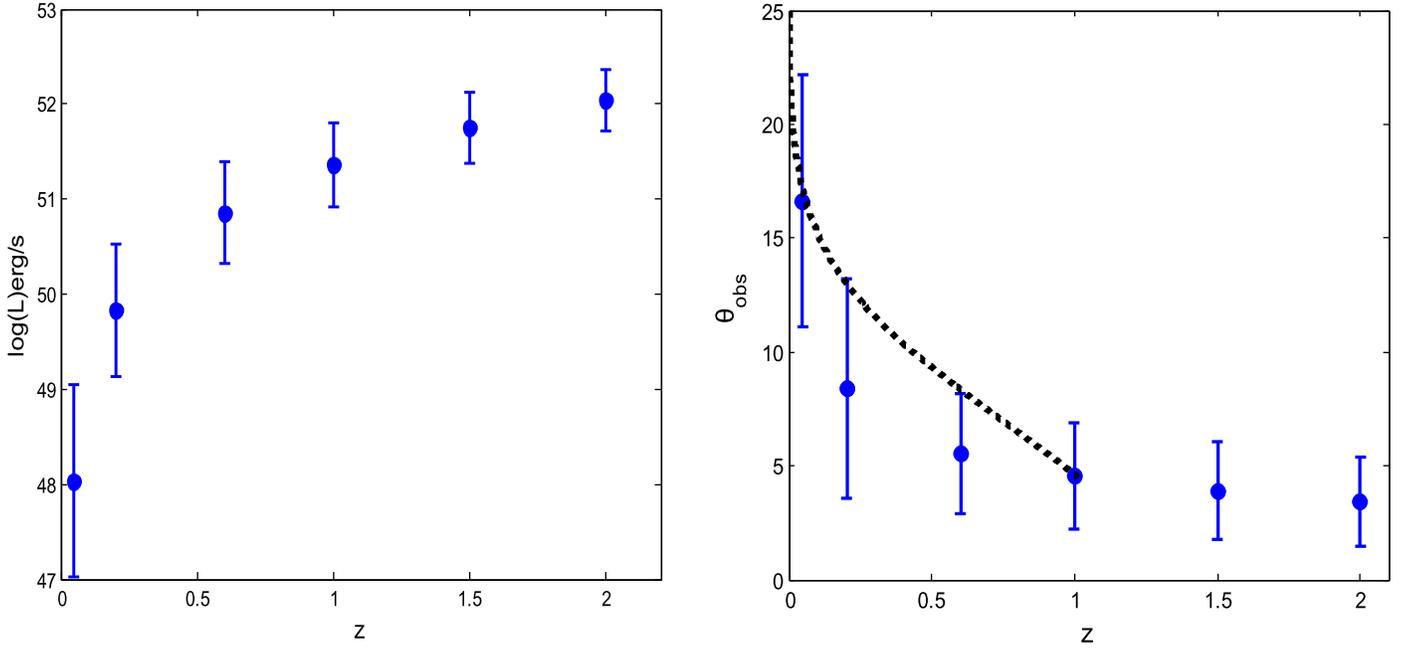

**Figure 5.** The evolution of the typical luminosity and viewing angle with redshift. The error bars are $1\sigma$ errors. The dashed line is the distribution of the maximum observable viewing angle obtained in Howell et al. (2019).

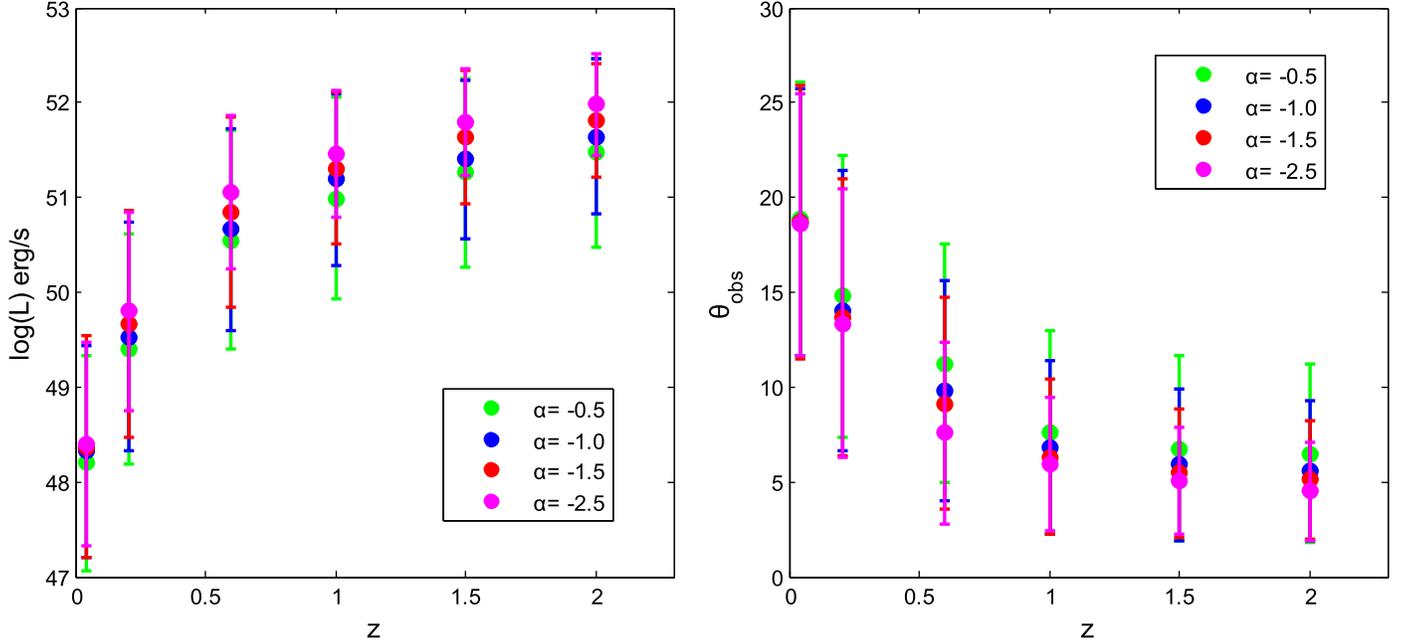

**Figure 6.** The evolution of typical luminosity and viewing angle of sGRBs within certain volumes with redshift. The power-law index $\alpha$ of the merger time distribution is taken to be $-0.5, -1.0, -1.5, -2.5$. The error bars are $1\sigma$ errors.

Here $f(E)$ is the Band function of sGRBs spectra (Band et al. 1993). $E_{\min}$ and $E_{\max}$ denote the lower and upper cutoff values of the detector's observational energy range, $\alpha_*$ and $\beta_*$ are the spectral indices, and we take the values $\alpha_* = -1$, $\beta_* = -2.5$, and $E_p = 800$ keV (Wanderman & Piran 2015; Lien et al. 2016; Mogushi et al. 2019).

## 3. Results of Monte Carlo Simulation

Based on the above model, we can derive the distribution of observed luminosity and the viewing angle $\theta_{\rm obs}$ of sGRBs with the Monte Carlo simulations. By assuming the uniform distribution of the solid angle $\Omega(\theta)$ in the sky, we can first calculate the Lorentz factor and the luminosity of sGRBs with Equations (1)–(5). Here we assume that $L_0$ satisfies the distribution of Equation (9) (Wanderman & Piran 2015) and take $\Gamma_0 = 300$ (Beniamini & Nakar 2019). Several papers have investigated the emission of GRB 170817A with the structured jet model and found that $\theta_0 \sim 5°$ can reproduce the observed characteristics well (e.g., Troja et al. 2019), so here we take $\theta_0 = 5°$. From Equation (6), we can obtain the redshift distribution of sGRBs. To obtain the observed luminosity and viewing angle $\theta_{\rm obs}$ distribution, the selection effect must be considered. Because the Fermi-GBM telescope has a larger field of view (Burns et al. 2016), we take the Fermi-GBM 64 ms limiting flux ($F_{\rm lim} = 2.0 \times 10^{-7}$ erg cm$^{-2}$ s$^{-1}$) as the detection threshold (Goldstein et al. 2017). The sGRB can be detected only when its flux is larger than this threshold.

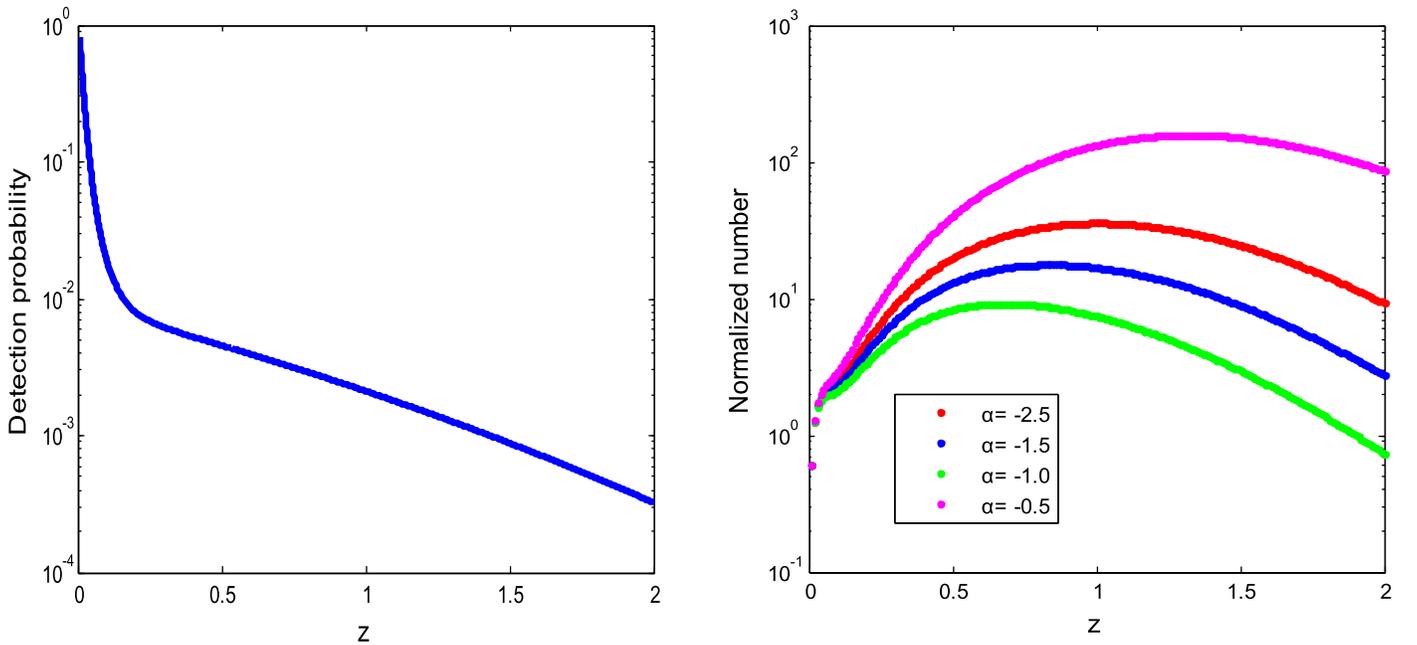

**Figure 7.** Left is the distribution of detection probability with redshift. Right is the normalized number distribution of observed sGRBs with our simulation, where the dotted lines are the 1σ uncertainty range. The green, blue, red, and purple lines correspond to different values of α with α = −0.5, −1.0, −1.5, −2.5.

The distribution of observed gamma-ray luminosity and viewing angle at different redshifts are shown in Figures 3 and 4. It can be seen that, for lower redshifts, the typical luminosity is relatively smaller, while the typical viewing angle is larger. For example, at $z = 0.045$ and $0.208$ (corresponding to the luminosity distance of 200 Mpc and 1 Gpc), the peak luminosities are about $8 \times 10^{47}$ erg s$^{-1}$ and $3 \times 10^{49}$ erg s$^{-1}$, respectively, while the peak viewing angles are around 17° and 8°. Therefore, the luminosity/viewing-angle distribution for nearby sGRBs is quite different from previous thinking. Figure 5 gives the variation of the typical luminosity and viewing angle with redshift. It is obvious that the observed typical luminosity increases with redshift, while the typical viewing angle decreases with redshift. Especially at smaller redshifts, the luminosity rises rapidly while, at larger redshifts (such as $z > 1$), they increase slowly. This is because, for GRBs with larger redshifts, only an observer with smaller viewing angles (for $z > 1$ the viewing angle should close to the energetic core of the jet) can detect them under the structured jet scenario (see also Gupte & Bartos 2018; Howell et al. 2019). We note that our results are consistent with the distribution of the maximum viewing angles obtained by Howell et al. (2019).

Figure 6 shows the luminosity and viewing-angle distributions of sGRBs within certain volumes. Since the sensitive distance of advanced LIGO/Virgo is about 200 Mpc for binary neutron star mergers or 400 Mpc for typical neutron star–black hole mergers (Abadie et al. 2010), so the sGRBs associated with present GW events are all local, at smaller redshifts. From these figures, we can see that, for sGRBs from binary neutron star mergers detected by advanced LIGO/Virgo, their luminosity mainly lie between $10^{47}$ and $10^{48}$ erg s$^{-1}$. For the third generation GW detector (i.e., Einstein Telescope), the redshifts of GW events can reach to ∼1.[3] In this case, the typical luminosity of sGRBs are $8 \times 10^{50}$–$7 \times 10^{51}$ erg s$^{-1}$.

We also calculate these distributions with α between −0.5 and −2.5 to see whether different models of formation rate of sGRBs will affect these distributions significantly. Figure 6 shows the variation of the typical luminosity and viewing angle with redshift for four different values of α. As can be seen from the figure, the typical luminosity of sGRBs increases very slightly with the decrease of α and, in fact, the discrepancies between these results are so small that it is hard to discriminate between different models. This is also true for the viewing-angle distribution, although the typical viewing angle is somewhat smaller for smaller values of α, the difference is not significant, i.e., the distribution of viewing angles is also insensitive to the value of α. This indicates that the formation rate of sGRBs with different power-law time delay index α almost does not affect the observed luminosity and viewing-angle distributions.

By comparing the total simulated number of sGRBs with what can be detected by Fermi-GBM, we obtained the detection probability of sGRBs by Fermi-GBM; the results are shown in the left panel of Figure 7. We can see that the detection probability decreases with redshift; when the redshift is between 0 and 0.2, the detection probability drops very sharply and, at $z = 1$, the detection probability is only about $10^{-3}$. We also give the normalized distribution of number of observed sGRBs at different redshifts, as shown in the right panel of Figure 7. Here we have calculated the distribution of number of sGRBs for four different values of α and it is very interesting that these distributions are obviously different. For smaller values of α, the redshift corresponding to the peak number of observed sGRBs is larger, and at redshift $z = 1$, the observed number of sGRBs for α = −2.5 is much larger than that for $\alpha = -0.5, -1$, and $-1.5$, thus it provides a better way to determine the values of $\alpha$.

---
[3] http://www.et-gw.eu/index.php/etdsdocument

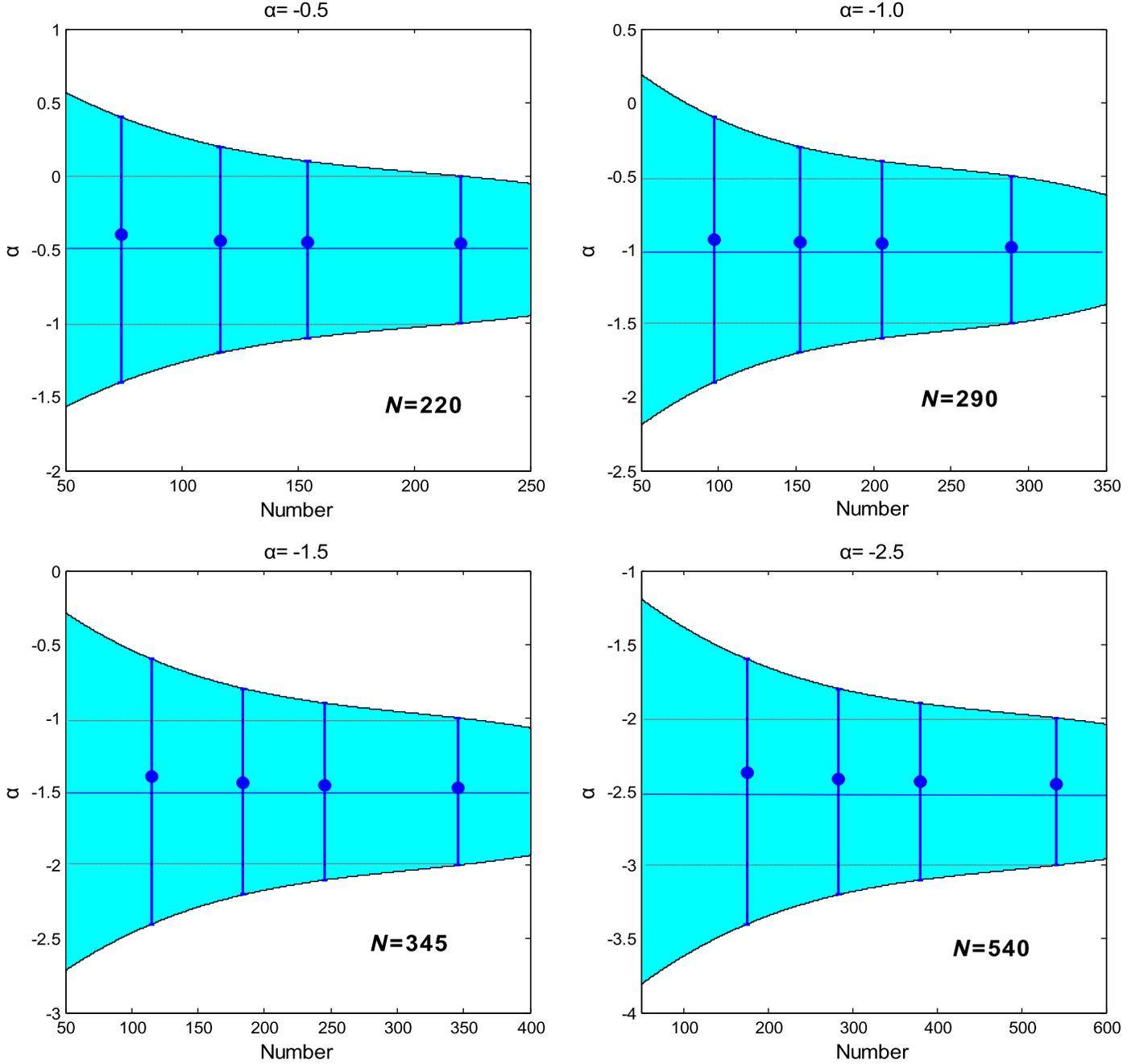

**Figure 8.** Evolution of the 95% confidence interval of $\alpha$ inferred from different simulated sample sizes. The black dotted line represents the true injected value and the red solid line is the two adjacent $\alpha$ values differing by 0.5. The black solid lines trace the averaged recovered confidence intervals, and the averaged most probable $\alpha$ values are marked with blue dots.

We use Bayesian method to estimate the number of sGRBs required to distinguish different $\alpha$ values. The posterior of $\alpha$ after observing a series of events with redshift can be obtained under the standard Bayesian framework:

$$P(\hat{\alpha}|\,D) \propto L(D|\,\hat{\alpha})P(\hat{\alpha}) \quad (12)$$

$$L(D|\,\hat{\alpha}) = \prod_i^n (N_\alpha^*(z_i)). \quad (13)$$

In Equation (12), $L(D|\,\hat{\alpha})$ is the likelihood of the data series $D$ predicted by the theoretical distribution calculated from a specific $\alpha$ ($N_\alpha^*(z)$, as shown in Figure 7). $P(\hat{\alpha})$ is the prior, which is taken to be uniform in this study, and the likelihood term can be given by Equation (13). We draw random redshift samples with $z < 1$ from $N_\alpha^*(z)$ for different $\alpha$ values ($-0.5$, $-1$, $-1.5$, and $-2.5$), then recover the $\alpha$ values by calculating their posterior distributions described by Equation (12). The number of samples progressively increases, so that we can investigate how the constraint on $\alpha$ changes with the observed GRB numbers. This process is repeated 100 times for each injected $\alpha$, and the averaged 95% confidence intervals of the recovered $\alpha$ are shown in Figure 8. The number of samples needed to let the 95% confidence intervals lie between the two adjacent $\alpha$ with respect to the injected one are 220, 290, 345, and 540 for $\boldsymbol{\alpha} = -0.5, -1.0, -1.5$, and $-2.5$, respectively.

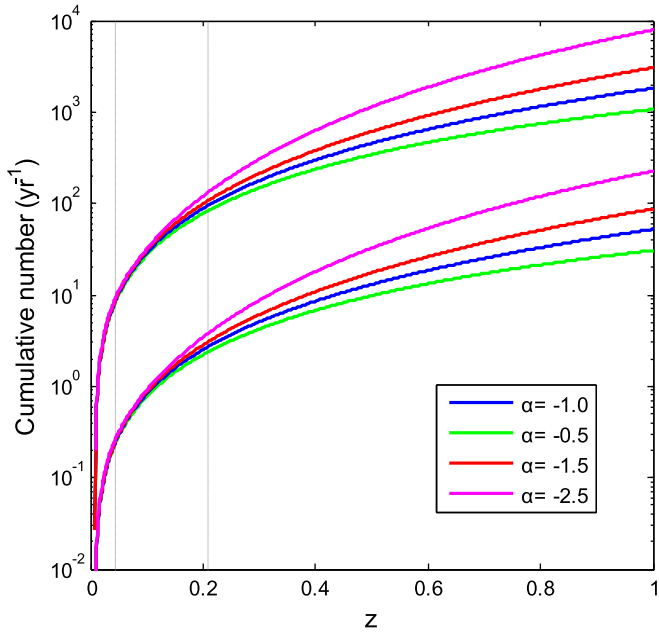

**Figure 9.** The expected detection rate of sGRBs with redshift. The solid and dotted lines represent the number of sGRBs with the rate of local NS–NS merger 110–3840 Gpc$^{-3}$ yr$^{-1}$ (Abbott et al. 2019). The two red lines represent the distances of 200 Mpc and 1 Gpc, respectively.

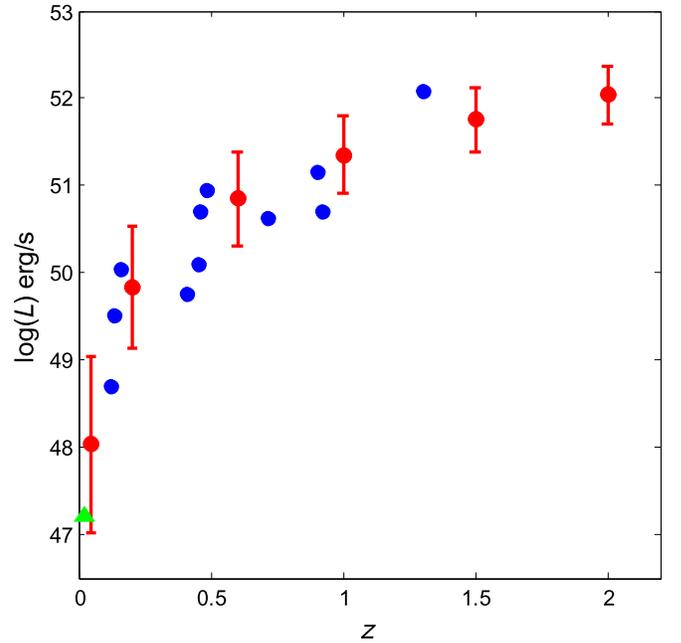

**Figure 10.** The relation between gamma-ray luminosity and redshift. The red circles represent our simulation results, the green triangle represents sGRB 170817A, and the blue circles are the sGRBs with known redshifts observed by Fermi-GBM (in Table 2). The error bars are 1$\sigma$ errors.

**Table 1**
The Expected Detection Rate of sGRBs within Some Redshifts

| Model | Redshift (distance) | The Expected Detection Rate (yr$^{-1}$) |
|---|---|---|
| $\alpha = -0.5$ | 0.045(200 Mpc) | 0.24 ~ 8.56 |
| | 0.208(1 Gpc) | 2.35 ~ 82.06 |
| | 1 | 30.57 ~ 1069 |
| $\alpha = -01.0$ | 0.045(200 Mpc) | 0.25 ~ 8.82 |
| | 0.208(1 Gpc) | 2.67 ~ 93.36 |
| | 1 | 52.08 ~ 1882.7 |
| $\alpha = -01.5$ | 0.045(200 Mpc) | 0.26 ~ 9.06 |
| | 0.208(1 Gpc) | 3.01 ~ 105.36 |
| | 1 | 87.17 ~ 3050.6 |
| $\alpha = -02.5$ | 0.045(200 Mpc) | 0.27 ~ 9.48 |
| | 0.208(1 Gpc) | 3.70 ~ 129.65 |
| | 1 | 225.88 ~ 7904.7 |

**Table 2**
List of sGRBs Detected by Fermi-GBM

| sGRB | z | $L_{51}$ | References |
|---|---|---|---|
| 080905A | 0.122 | 0.00492 | 1 |
| 090510 | 0.903 | 1.43 | 1 |
| 100117A | 0.920 | 0.503 | 1 |
| 100206A | 0.407 | 0.0566 | 1 |
| 131004A | 0.717 | 0.424 | 1 |
| 100625A | 0.453 | 0.126 | 1 |
| 111117A | 1.3 | 12 | 2 |
| 160821B | 0.16 | 0.11 | 2 |
| 160624A | 0.483 | 0.88 | 2 |
| 150120A | 0.46 | 0.51 | 2 |
| 150101B | 0.1343 | 0.032 | 2 |

**Note.** The redshift and luminosity of sGRBs are taken from 1. Wang et al. (2019), 2. Mogushi et al. (2019).

Given the detection probability, we can calculate the expected detection rate of sGRBs by Fermi-GBM (see also Howell et al. 2019; Mogushi et al. 2019)

$$R_{sGRB,obs}(z) = R_{sGRB}(z) f_{sGRB}(z) \Lambda_{GRB}. \quad (14)$$

where $R_{sGRB}(z)$ is the formation rate of sGRBs as given by Equation (6) and $f_{sGRB}(z)$ is the detection probability as shown in Figure 7. We use the total time-averaged observable sky fraction of Fermi-GBM which is given as $\Lambda_{GRB} = 0.60$ (Burns et al. 2016). In the top-hat model, the conversion of the observed rate to the intrinsic rate typically needs a geometrical beaming factor $f_b$, and $f_b = 1 - \cos(\theta_j) \approx \theta_j^2/2$, where $\theta_j$ is the jet half-opening angle. However, for the structure jet model, the effect of beaming factor has actually been included in the detection probability. Then we use Equation (14) to calculate the expected detection rate of sGRBs by Fermi-GBM. The results are shown in Figure 9, where we use the local rates of 110–3840 Gpc$^{-3}$ yr$^{-1}$ (Abbott et al. 2019) for binary neutron stars mergers and assume $\alpha = -0.5 \sim -2.5$. The expected detection rate of sGRBs within some distances have also been listed in Table 1. As can be seen from them, the change of $\alpha$ has little effect on the observed number of sGRBs at low redshift; this is not surprising since, from Figure 2, we can see that the changing $\alpha$ has negligible effect on the formation rate of sGRBs at low redshift. The detection rate of sGRBs by Fermi-GBM is about $39.65^{+71.98}_{-30.92}$ yr$^{-1}$ (Howell et al. 2019). By comparing this with our results, we find that the smaller value of $\alpha$ (such as $\alpha = -2.5$) is disfavored.

We also compare our results of gamma-ray luminosity distribution with sGRBs with known redshifts detected by Fermi-GBM, as shown in Figure 10 (the data is in Table 2). It is evident that our results are consistent with the observation. In Figure 11, We perform the Kolmogorov–Smirnov test for the cumulative luminosity distribution of our simulated sGRBs and those with known redshifts. The chance probabilities is 0.48, which indicates that our simulation results can reproduce the observed luminosity distribution well.

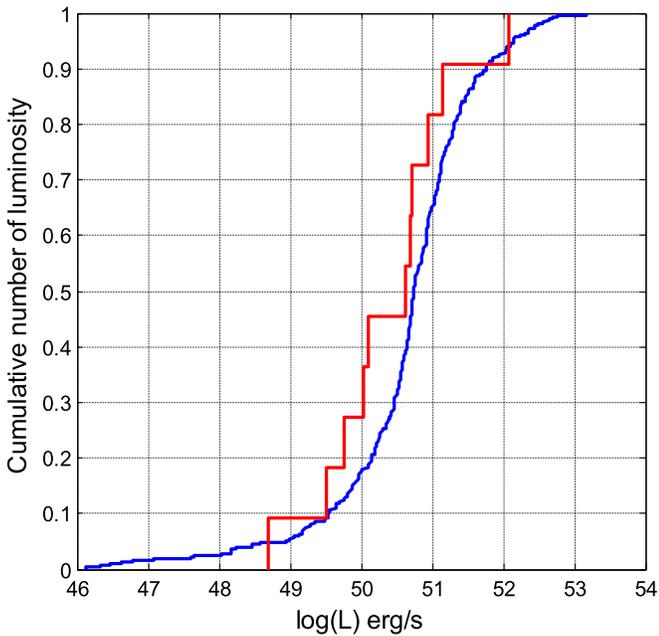

**Figure 11.** The cumulative luminosity distribution of our simulated sGRBs (blue line) and those with known redshifts (red line). The chance probabilities is 0.48.

## 4. Discussion and Conclusion

The joint detection of GW and gamma-ray emission from the binary neutron star merger event GW170817/GRB 170817A marks a breakthrough in the field of multi-messenger astronomy and confirms the association of sGRBs with binary neutron star mergers. The peculiar behavior of the prompt and afterglow emission of GRB 170817A indicates that this nearby sGRB was seen from a structured jet with large viewing angle, and this speculation has been identified by the successful measurement of the superluminal movement of the radio afterglow emission (Abbott et al. 2017a, 2017b; Goldstein et al. 2017; Savchenko et al. 2017).

Since the beginning of the third observation run of advanced LIGO/Virgo, a number of NS–NS and NS–BH merger events/candidates have been detected, such as the notable S190425z (LIGO Scientific Collaboration & Virgo Collaboration 2019a, 2019b) and S190426c (LIGO Scientific Collaboration & Virgo Collaboration 2019c). Although no sGRBs associated with them have been detected, it is consistent with our prediction about the expected detection rate of nearby sGRBs (see Figure 9). The detection of these NS–NS and NS–BH merger events imply that a number of potential sGRBs really exist within the sensitive distance of advanced LGO/Virgo and they could probably be detected by the next generation of more sensitive gamma-ray detectors, so it is important to investigate the properties of these nearby sGRBs. In this paper, we use Monte Carlo simulation to calculate the gamma-ray luminosity and viewing-angle distribution of nearby sGRBs based on the structured jet model, and obtain the evolution of these distributions with redshift. We find that, for sGRBs from binary neutron star mergers detected by advanced LIGO/Virgo, their luminosity mainly lies between $10^{47}$ and $10^{48}$ erg s$^{-1}$.

In our calculation, we assume the formation rate of sGRBs is proportional to the SFR. Although the SFR is uncertain at high redshifts, it is well known at low redshifts, and various SFR models are only slightly different at low-redshift region $z < 1.5$, so for the nearby sGRBs that we are most interested in, the uncertainty of SFR have nearly no effects on our results. Here we use the latest SFR model of Madau & Dickinson (2014) to calculate the sGRB rate. In addition to the SFR, the merger time distribution is also essential to estimate the sGRB rate. From Figure 6, we see that the typical luminosity and viewing angles of sGRBs seem to be insensitive to the value of $\alpha$, which means that we cannot identify the value of $\alpha$ through observed luminosity and viewing-angle distribution. However, it is very interesting that, for different values of $\alpha$, the distribution of numbers of observed sGRBs are quite different, so it is possible to determine the value of $\alpha$ through observed distribution of the number of sGRBs. We used the Bayesian method to make a quantitative analysis and found that the value of $\alpha$ may be identified when the number of observed sGRBs with known redshifts is more than 200.

During the advanced LIGO/Virgo O3 run a number of NS–NS and NS–BH merger events/candidates have been detected, but no corresponding sGRBs have been observed. This is consistent with our results that the expected detection rate of sGRBs is only about 1 yr$^{-1}$ by Fermi-GBM within the distance of several hundred Mpc, so developing more sensitive GRB detectors is very important for future studies of GRBs, especially for nearby sGRBs that may be weak and associated with GW events. The GECAM (Gravitational Wave Electromagnetic Counterpart All-sky Monitor) mission is the planned Chinese space telescope to monitor the GRBs coincident with GW events, which will launch in late 2020. It has a FOV of 100% all-sky and it has lower detection threshold than Fermi-GBM, about the same as Swift (Zhang et al. 2019; Zheng 2019). Our calculation shows that the peak luminosity is about $6 \times 10^{46}$ erg s$^{-1}$ and $4.0 \times 10^{48}$ erg s$^{-1}$ at luminosity distance of 200 Mpc and 1 Gpc, respectively, and the number of sGRBs that can be detected by GECAM are 1.04−35.96 yr$^{-1}$ and 16.97−593.91 yr$^{-1}$ within 200 Mpc and 1Gpc, respectively, which is about five times larger than that can be detected by Fermi-GBM.

We are grateful to the anonymous referee for valuable comments. This work was supported by NSFC (No. 11921003 and No. 11933010), by the Chinese Academy of Sciences via the Strategic Priority Research Program (No. XDB23040000), and Key Research Program of Frontier Sciences (No. QYZDJ-SSW-SYS024).

ORCID iDs

Qi Guo https://orcid.org/0000-0001-9048-4616
Daming Wei https://orcid.org/0000-0002-9758-5476
Yuanzhu Wang https://orcid.org/0000-0001-9626-9319

References

Abadie, J., Abbott, B. P., Abbott, R., et al. 2010, CQGra, 27, 173001
Abbott, B. P., Abbott, R., Abbott, T. D., et al. 2017a, PhRvL, 119, 161101
Abbott, B. P., Abbott, R., Abbott, T. D., et al. 2017b, ApJL, 848, L12
Abbott, B. P., Abbott, R., Abbott, T. D., et al. 2019, PhRvX, 9, 031040
Aloy, M. A., Janka, H.-T., & Muller, E. 2005, A&A, 436, 273

Ando, S. 2004, JCAP, 06, 007
Band, D., Matteson, J., Ford, L., et al. 1993, ApJ, 413, 281
Belczynski, K., Bulik, T., Olejak, A., et al. 2018, arXiv:1812.10065
Belczynski, K., Klencki, J., Fields, C. E., et al. 2017, arXiv:1706.07053
Beniamini, P., & Nakar, E. 2019, MNRAS, 482, 5430
Berger, E. 2014, ARA&A, 52, 43
Berger, E., Fong, W., & Chornock, R. 2013, ApJL, 744, L23
Berger, E., Kulkarni, S. R., Pooley, G., et al. 2003, Natur, 426, 154
Bloom, J. S., Frail, D. A., & Sari, R. 2001, AJ, 112, 2879
Bulik, T., Belczynski, K., & Rudak, B. 2004, A&A, 415, 407
Burns, E., Connaughton, V., Zhang, B.-B., et al. 2016, ApJ, 818, 110
Dai, Z. G., & Gou, L. J. 2001, ApJ, 552, 72
Gehrels, N., Ramirez-Ruiz, E., & Fox, D. B. 2009, ARA&A, 47, 567
Ghirlanda, G., Ghisellini, G., & Celotti, A. 2004, A&A, 422, L55
Ghirlanda, G., Salafia, O. S., Paragi, Z., et al. 2019, Sci, 363, 968
Goldstein, A., Veres, P., Burns, E., et al. 2017, ApJL, 848, L14
Granot, J., Panaitescu, A., Kumar, P., & Woosley, S. E. 2002, ApJL, 570, L61
Gupte, N., & Bartos, I. 2018, arXiv:1808.06238
Howell, E., Ackley, K., Rowlinson, A., & Coward, D. 2019, MNRAS, 485, 1435
Huang, Y. F., Cheng, K. S., & Gao, T. T. 2006, ApJ, 637, 873
Jin, Z.-P., Covino, S., Liao, N.-H., et al. 2020, NatAs, 4, 77
Jin, Z.-P., Hotokezaka, K., Li, X., et al. 2016, NatCo, 7, 12898
Jin, Z.-P., Li, X., Cano, Z., et al. 2015, ApJL, 811, L22
Jin, Z.-P., Yan, T., Fan, Y. Z., & Wei, D. M. 2007, ApJL, 656, L57
Kathirgamaraju, A., Barniol Duran, R., & Giannios, D. 2018, MNRAS, 473, L121
Lazzati, D., Perna, R., Morsony, B. J., et al. 2018, PhRvL, 120, 241103
Lien, A., Sakamoto, T., Barthelmy, S. D., et al. 2016, ApJ, 829, 7
LIGO Scientific Collaboration & Virgo Collaboration 2019a, GCN, 24168
LIGO Scientific Collaboration & Virgo Collaboration 2019b, GCN, 24228
LIGO Scientific Collaboration & Virgo Collaboration 2019c, GCN, 24237
Madau, P., & Dickinson, M. 2014, ARA&A, 52, 415M
Meacher, D., Coughlin, M., Morris, S., et al. 2015, PhRvD, 92, 063002
Mészáros, P. 2006, RPPh, 69, 2259
Mogushi, K., Cavaglia, M., & Siellez, K. 2019, ApJ, 880, 55
Mooley, K. P., Deller, A. T., Gottlieb, O., et al. 2018, Natur, 561, 355
Murguia-Berthier, A., Ramirez-Ruiz, E., Montes, G., et al. 2017, ApJL, 835, L34
Nakar, E. 2007, PhR, 442, 166
Pian, E., D'Avanzo, P., Benetti, S., et al. 2017, Natur, 551, 67
Porciani, C., & Madau, P. 2001, ApJ, 548, 522
Regimbau, T., & Hughes, S. A. 2009, PhRvD, 79, 062002
Rossi, E., Lazzati, D., & Rees, M. J. 2002, MNRAS, 332, 945
Savchenko, V., Ferrigno, C., Kuulkers, E., et al. 2017, ApJL, 848, L15
Schmidt, M. 2001a, ApJ, 552, 36
Schmidt, M. 2001b, ApJL, 559, L79
Tanvir, N. R., Levan, A. J., Fruchter, A. S., et al. 2013, Natur, 500, 547
Troja, E., van Eerten, H., Ryan, G., et al. 2019, MNRAS, 489, 1919
Wanderman, D., & Piran, T. 2015, MNRAS, 448, 3026
Wang, F., Zou, Y.-C., Liu, F., et al. 2019, arXiv:1902.05489
Wei, D. M., & Jin, Z. P. 2003, A&A, 400, 415
Woosley, S. E. 1993, ApJ, 405, 273
Wu, X. F., Dai, Z. G., Huang, Y. F., & Lu, T. 2005, MNRAS, 357, 1197
Yang, B., Jin, Z. P., Li, X., et al. 2015, NatCo, 6, 7323
Zhang, B. 2007, ChJAA, 7, 1
Zhang, B., Dai, X., Lloyd-Ronning, N. M., & Mészáros, P. 2004, ApJL, 601, L119
Zhang, B., & Mészáros, P. 2002, ApJ, 571, 876
Zhang, B., & Mészáros, P. 2004, IJMPA, 19, 2385
Zhang, D., Li, X., Xiong, S., et al. 2019, NIMPA, 921, 8
Zheng, S. 2019, in Proc. Extragalactic Explosive Universe 2019 Conf., The New Era of Transi-ent Surveys and Data-driven Discovery (Garching: ESO), 63